\documentstyle[12pt]{article}
\def\gdot#1{\rlap{$#1$}/}
\begin{document}
\begin{center}
{\LARGE{\bf THE AXIAL CHARGE\\

\vspace{0,3cm}

RENORMALIZATION IN A\\

\vspace{0,3cm}

RELATIVISTIC DESCRIPTION\\

\vspace{0,7cm}

OF FINITE NUCLEI}}

\end{center}

\begin{center}

{\large{A.~Gil, M.~Kleinmann, H.~M\"uther and E.~Oset}}

\end{center}

\begin{center}

{\small{\it Departamento de F\'{\i}sica Te\'orica and IFIC\\
Centro Mixto Universidad de Valencia - CSIC\\
46100 Burjassot (Valencia), Spain\\
and\\
Institut f\"ur Theoretische Physik\\
Universit\"at T\"ubingen\\
72076 T\"ubingen, Germany}}

\end{center}

\vspace{3cm}

\begin{abstract}

{\small{Starting from a realistic One-Boson-Exchange model of the
nucleon nucleon interaction the relativistic mean field for nucleons is
determined within the Dirac Brueckner Hartree Fock approach for finite
nuclei. The matrix elements of the axial charge operator evaluated for
the solutions of the Dirac equation with this selfenergy are
investigated. These matrix elements are enhanced with
respect to the equivalent non relativistic ones obtained from the solutions
of the Schr\"odinger equation with the non relativistic equivalent potential.
The
present results confirm at a qualitative level the results for the axial
charge renormalization obtained with perturbative approaches. However, the
results obtained differ in size from those of the perturbative approach and
are nucleus and state dependent.}}

\end{abstract}

\newpage

\section{Introduction.}

As suggested in ref. \cite{1} the presence of a large scalar potential in a
relativistic version of the nucleon selfenergy in the nucleus \cite{2} leads
to a sizeable renormalization of the axial charge in nuclei. This
renormalization, which is also sometimes refered to as the heavy meson
exchange current contribution,   must be considered in addition to the
conventional  meson exchange currents studied earlier \cite{3,4,5,6}.
More quantitative
evaluations of this renormalization, following the idea of \cite{1}, have been
recently provided in \cite{7,7a,8}. In ref. \cite{7} a perturbative
approach is
used starting from a relativistic description of the $NN$ potential and taking
direct and exchange terms. The strong short-range and tensor components of a
realistic $NN$ interaction give rize to significant two-nucleon
correlations. The effects of $NN$ correlations are taken into account in
the investigations of ref. \cite{7a} by using the Brueckner G-matrix.
The estimates reported in \cite{7} and \cite{7a} were made for the
system of infinite nuclear matter.

The investigations of ref. \cite{8} are performed directly for
finite nuclei. Also in this case the effect of the nucleon selfenergy is
treated in a perturbative way. The operators are reduced to a bispinor
representation and the calculations are carried out
in a nonrelativistic frame. The single-particle wavefunctions are
represented by oscillator wavefunctions and the effect of correlations
are included in terms of a simple local correlation function.

In the present work we want to consider the relativistic features, the
effects of correlations and the single-particle wavefunctions
consistently. For that purpose we employ the results of the
relativistic Dirac Brueckner Hartree Fock (DBHF) calculations of ref.
\cite{10}. These calculations are based on the version $A$ of the
relativistic One-Boson-Exchange potential of \cite{rupr}. The results
of the calculation of the ground-state properties of double
closed-shell nuclei are in good agreement with the experimental data
and the resulting self-energy yields a real part for the optical
potential of low-energy nucleon nucleus scattering, which is close to
the empirical analysis \cite{klein}.

After this short introduction we will review the perturbative treatment
of the heavy meson exchange current contribution in nuclear matter. The
self-consistent DBHF calculations are discussed in section 3, while
section 4 contains a discussion of the non-relativistic reduction. The
results are presented and discussed in section 5 and the final section
summarizes the main conclusions.

\section{Perturbative renormalization of the axial charge in a relativistic
approach.}

A realistic $NN$ interaction contains a large attractive scalar
isoscalar component (due to $\sigma$-exchange in OBE model) and a
repulsive vector isovector component ($\omega$-exchange). Evaluating
the selfenergy of a nucleon in a medium of nuclear matter with such an
interaction using the mean field approximation, one finds that it
contains a large attractive scalar component and a repulsive component,
which under Lorentz transformation transforms like the timelike
component of a vector \cite{2}

\begin{equation}
\Sigma  = \Sigma^{s}\frac{ \rho }{\rho_{0}} + \Sigma ^{v}\gamma^{0}
\frac{ \rho }{\rho_{0}} \label{self}
\end{equation}

\noindent
with $\rho$ the nuclear density and $\rho_{0}$ the saturation density
of nuclear matter $(\rho_{0} = 0.17 fm^{-3})$. Taking into account the
Fock-exchange terms in the Hartree-Fock approximation or accounting for
correlation effects in the DBHF approximation one obtains a small
spacelike vector component and finds that all the terms depend slightly
on the momentum of the nucleon \cite{10}. We now want to calculate matrix
elements for the axial operator $g_{A} \gamma^{\mu} \gamma_{5}$,
concentrating on the axial charge $g_{A} \gamma^{0} \gamma_{5}$ for
nucleons moving in the nuclear medium.
We can immediately write the perturbative corrections to the axial charge
due to the nucleon selfenergy, which are depicted diagramatically in fig.
1,
where we have separated the contribution from positive and negative
intermediate states in the nucleon propagator. Analytically this decomposition
is given by

\begin{equation}
\frac{\gdot{p} + M}{p^{2} - M^{2}} = \frac{M}{E(\vec{p}\:)} \{
\frac{\sum_{r} u_{r}(\vec{p}\:) \bar{u}_{r} (\vec{p}\:)}{p^{0} - E(\vec{p}\:)
+ i \epsilon} + \frac{\sum_{r} v_{r}(-\vec{p}\:) \bar{v}_{r}(-\vec{p}\:)}
{p^{0} + E (\vec{p}\:) - i \epsilon} \}
\end{equation}

\noindent
where $M, E (\vec{p}\:)$ are the mass and on shell energy of the free
nucleon and $u_{r}, v_{r}$ the ordinary free spinors in Mandl-Shaw
representation \cite{9}.
The axial charge matrix element is reduced to a bispinor representation
assuming $E(\vec{p}\:) \simeq M$ by means of

\begin{equation}
\bar{u} (\vec{p}\:') \gamma^{0} \gamma_{5} u(\vec{p}) =
\chi' \frac{\vec{\sigma}(\vec{p}
+ \vec{p}\:')}{2M} \chi \label{bare}
\end{equation}

Now the a) and b) diagrams from fig. 1 with positive intermediate nucleon
components are automatically absorbed into the calculation with dressed non
relativistic wave functions but genuine corrections from the negative
intermediate states c) and d) remain. One can easily see that the
renormalization with the $\Sigma ^{v}\gamma^{0}$ term of (\ref{self})
vanishes identically
and only the renormalization with the $\Sigma^{s}$ term remains. One
immediately gets a renormalized axial charge matrix element corresponding to
bare matrix element plus figs. 1c and 1d given by

\begin{equation}
g_{A} (1- \frac{\Sigma^{s}}{M} \frac{\rho}{\rho_{0}}) \chi' \frac{\vec{\sigma}
(\vec{p} + \vec{p}\:')}{2M} \chi \label{pertur1}
\end{equation}

\noindent
or equivalently a renormalization of the axial charge by the amount
\mbox{
$(1- \frac{\Sigma^{s}}{M} \frac{\rho}{\rho_{0}})$}.
This is the result
obtained in
\cite{7}.
Note that since the relativistic potential of (1) implicitly
accounts for direct and exchange terms no further corrections have to be done
in contrast to \cite{7} where, because one starts from a $NN$ interaction,
direct and exchange terms are explicitly evaluated. With standard values of
$\Sigma^{s}$ of the order of $-400 MeV$ and taking $\rho \simeq \rho_{0}$ one
obtains a renormalization factor of the order of 1.4 in qualitative agreement
with \cite{7,8}.

Another way to arrive at eq.(\ref{pertur1}) is to realize that the
solution of the Dirac equation for with a selfenergy of the kind
displayed in eq.(\ref{self}) yields Dirac spinors for the nucleons in
the nuclear medium, which are identical to Dirac spinors of free
nucleons, except that the mass of the nucleon $M$ has to be replaced by
an effective mass $M^*=M+\Sigma^{s}\rho / \rho_{0}$. Calculating the
matrix element for the axial charge operator with these dressed Dirac
spinors and reducing it to a bispinor representation one finds as in
eq.(\ref{bare})

\begin{eqnarray}
\bar{\tilde u} (p') \gamma^{0} \gamma_{5} \tilde u(p) & = &\chi'
\frac{\vec{\sigma}(\vec{p} + \vec{p}\:')}{2\left(M+\Sigma^{s}\frac{\rho}{
\rho_{0}}\right) } \chi \nonumber\\
& =  &\left[1- \frac{\Sigma^{s}}{M} \frac{\rho}{\rho_{0}}+ \left(
\frac{\Sigma^{s}}{M} \frac{\rho}{\rho_{0}}\right)^2 + \cdots \right]
\chi' \frac{\vec{\sigma} (\vec{p} + \vec{p}\:')}{2M} \chi
\label{pertur2}
\end{eqnarray}
It should be noted that this non-perturbative treatment of the
the heavy meson exchange current contribution to the renormalization of
the axial charge yields an effect which is considerably larger than
the perturbative treatment of eq.(\ref{pertur1}). Using again $\Sigma^{s}
 = -400 MeV$ and taking $\rho \simeq \rho_{0}$ one obtains a factor of
1.7 rather than 1.4 (see above).

\section{Finite nuclei renormalization.}

Solving the Dirac equation directly for the finite nucleus the diagrams
in fig. 1 plus all terms of higher order in the nucleon selfenergy insertions
can automatically be taken into account by
evaluating the matrix elements of the $\gamma^{0} \gamma_{5}$ operator between
the solutions of the Dirac equation. The relativistic selfenergy for
the finite nucleus is calculated following the scheme defined as
Hartree approximation in ref.~\cite{10}. In this scheme we assume an
effective Lagrangian for nucleons, a scalar $\sigma$ meson and a vector
$\omega$ meson. This effective Lagrangian is defined to be used in
Dirac Hartree calculations for finite nuclei. The effects of
correlations and the Fock-exchange terms are taken into account
by assuming coupling constants for the meson nucleon interaction terms,
which are density dependent and which are determined such that a Dirac Hartree
calculation of nuclear matter reproduces the results of microscopic
DBHF calculations for the OBE potential $A$ at all densities.

In this scheme the correlation effects are deduced from nuclear matter
and treated in a local density approximation. The investigations of
ref.\cite{10} demonstrate that these Dirac Hartree calculations yield
results which are close to DBHF calculations in which the correlation
effects are treated directly for the finite nucleus under
consideration. Furthermore the predictions of this Dirac Hartree
approximation for binding energy and radius of nuclei like $^{16}$O and
$^{40}$Ca are close to the empirical data. Furthermore it is worth
noting that this approach leads to scalar and vector components of the
self-energy

\begin{equation}
\Sigma (r)= \Sigma ^{s} (r) + \gamma^{0} \Sigma^{v} (r) \label{self2}
\end{equation}

\noindent
which are local and depend on the radial distance $r$ only. Using this
selfenergy in the Dirac equation

\begin{equation}
[i \vec{\gamma} \vec{\nabla} + M + \Sigma ^{s} (r) + \gamma^{0} \Sigma
^{v}(r)] \Psi = \gamma^{0} E \Psi
\end{equation}

\noindent
we obtain solutions for the energy $E$ and Dirac spinors with the
various quantum numbers for the orbital angular momentum $l$ and total
angular momentum $j$. By following the nomenclature of Itzykson and
Zuber \cite{itz} we write the relativistic wave functions as

\begin{equation}
\Psi_{jm} ^{l}(\vec{r}) \equiv
\left\{
\begin{array}{ll}
i \frac{G_{lj}(r)}{r} & \psi_{jm} ^{l}\\
\\
\frac{F_{lj}(r)}{r} & \vec{\sigma} . \hat{r} \psi_{jm} ^{l}
\end{array}
\right\}
\end{equation}

\noindent
where $\psi_{jm}^{l}$ are the spin wave functions given by

$$
\psi_{jm}^{(+)}=(\frac{1}{2j})^{1/2}
\left\{
\begin{array}{ll}
(j + m)^{1/2} & Y_{j-1/2}^{m-1/2}\\
(j - m)^{1/2} & Y_{j-1/2}^{m+1/2}
\end{array}
\right\}
for\;\: j = l + 1/2
$$

\begin{equation}
\psi_{jm}^{(-)} = (\frac{1}{2j+2})^{1/2}
\left\{
\begin{array}{ll}
(j+1-m)^{1/2} & Y_{j+1/2}^{m-1/2}\\
-(j+1+m)^{1/2} & Y_{j+1/2}^{m+1/2}
\end{array}
\right\}
for\;\: j=l-1/2\;\:(l>0)
\end{equation}

With a little bit of algebra the matrix elements of $\gamma^{0} \gamma_{5}$
between Dirac wave functions are readily evaluated and the results are shown
in the appendix.

Since the axial charge renormalization is checked in the $0^{+}\leftrightarrow
0^{-}$ first forbidden $\beta$-decay transitions we have performed the
calculations for the relevant matrix elements in these transitions in two
nuclei $^{16}O$ and $^{40}Ca$, in order to see the difference of the
renormalization for nuclei with different mass number.

For such double closed shell nuclei a $0^{-}$ state is formed with a $ph$
excitation for $l_{p}$ and $l_{h}$ with different
parity and $j_{p}=j_{h}$. We have

\begin{eqnarray}
| 0^{-} >&=&\sum_{m}(-1)^{j_{h}-m} C(j_{p} j_{h} 0; m, -m) a^\dagger_{j_{p}m}
a_{j_{h}m} | 0^+ >
\nonumber\\
&=& \sum_{m}\sqrt{\frac{1}{2j_{h}+1}}
a^\dagger_{j_{p}m} a_{j_{h}m} | 0^+ >
\end{eqnarray}

The matrix element for the $0^{+} \leftrightarrow 0^{-}$ transition is readily
obtained from the formulas in the appendix by setting $j = j', m = m'$
summing over $m$ and multiplying by $(2j + 1)^{-1/2}$. The $0^{+}
\leftrightarrow 0^{-}$ transition can only be done with cases b) and c) and
in both cases we get a simplified solution which is

$$
< 0^{+} | \gamma^{0} \gamma_{5} e^{i \vec{q} \vec{r}} | 0^{-} > =
$$

\begin{equation}
\sqrt{2j + 1} (-i) \int r^{2} dr \left[ \frac{G_{l' j'} (r)}{r} \frac{F_{lj}
(r)}
{r} - \frac{F_{l' j'}(r)}{r} \frac{G_{lj} (r)}{r} \right] j_{0} (qr)
\end{equation}

We construct the $0^{-}$ state in $^{16}O$ as a $ph$ excitation with the
orbitals $1p_{1/2}$ and $2s_{1/2}$. We also consider the component with
$1p_{3/2}$ and $1d_{3/2}$ in order to see whether the renormalization is
state dependent or not. In the case of $^{40}Ca$ we take the orbitals
$1d_{3/2}$ and $2p_{3/2}$ . In all cases we consider a proton hole
state and a neutron particle state, as
it corresponds to $\beta$ transitions. The neutron states in
the $2s_{1/2}$ and $1d_{3/2}$ orbitals in $^{16}O$ and the $2p_{3/2}$ orbital
of $^{40}Ca$ are all bound states in the potential used.

\section{Non relativistic calculation.}

In order to see the effects of the renormalization due to the relativistic
structure of the potential (\ref{self2}) we solve the Schr\"odinger
equation with the equivalent nonrelativistic potential
\cite{jam89,klein}

\begin{equation}
U_{SEP}(r) = \Sigma^s(r) + \frac{E}{M}\Sigma^v(r) +
\frac{\left(\Sigma^s(r)\right)^2 - \left(\Sigma^v(r)\right)^2 +
U_{\hbox{Darwin}}(r)}{2M} \label{eq:sep}
\end{equation}
with
\begin{eqnarray}
U_{\hbox{Darwin}}(r) & = & \frac{3}{4} \left[
\frac{1}{D(r)}\frac{d\,D(r)}{dr}\right]^2 - \frac{1}{rD(r)}\frac{d\,D(r)}{dr}
-\frac{1}{2D(r)} \frac{d^2D(r)}{dr^2}\; , \nonumber\\
D(r) & = &  E + \Sigma^s(r) - \Sigma^v(r) \; .
\end{eqnarray}

\noindent
The single-particle wavefunctions obtained from the solution of the
Schroedinger equation with $U_{SEP}$ are used to evaluate the matrix
elements of the $\vec{\sigma} (\vec{p} + \vec{p}\:')/2M$ operator and
we find

$$
< 0^{+} | \vec{\sigma} \frac{(\vec{p} + \vec{p}')}{2M} | 0^{-} >=
$$

\begin{equation}
i(-1)^{j' + l' + 1/2} \frac{1}{2M} \sqrt{2} F (n' l' j', nlj; \lambda = 0)
\end{equation}

\noindent
where the function $F$ is defined in the appendix.

\section{Results and discussion.}

In fig. 2 we show the matrix elements of the axial charge for the relativistic
and non relativistic cases of eqs. (11) and (14) respectively as a
function of $q$. We show the results for the $1p_{1/2} \rightarrow 2s_{1/2}$
and $1p_{3/2} \rightarrow 1d_{3/2}$ transitions on $^{16}O$. One
observes that the strength of the latter transition is about a factor 2
larger than for the first one. In both cases the relativistic
calculation of the matrix element yields larger values than the
non-relativistic one derived from the equivalent non relativistic potential.
We also observe that the matrix elements are weakly dependent on the momentum
transfer $q$ up to values of $q \simeq 100\:MeV/c$. In fig. 3 we
show the ratio of the relativistic versus non relativistic matrix elements as
a function of $q$ for the two transitions in $^{16}O$.
The values of the ratios at $q=0$, relevant to $\beta$ decay, are about 1.33
and 1.20 for the $1p_{1/2} \rightarrow 2s_{1/2}$ and $1p_{3/2}
\rightarrow 1d_{3/2}$ transitions, respectively.

The strength of $\Sigma ^{s}
(r\:)$ from eq. (6) at $r =0$ is in our case $\Sigma^{s} \simeq - 384\:
MeV$, close to the value typical for nuclear matter at saturation (
$\Sigma^{s} \simeq - 400\:  MeV$) which we have considered in our
estimates of section 2. In the perturbative approach of section 2 we would
have obtained a ratio of 1.41 whereas the non perturbative nuclear
matter estimate would even yield a ratio of 1.69 for this value of
$\Sigma^{s}$.  We can see that the calculations performed
directly for the finite nuclei yields results which are significantly
smaller than those estimates from nuclear matter. The reason for this
difference is the fact that the calculation of matrix elements for
finite nuclei requires a radial integration which is dominated by the
integrand at the surface. This is due to the fact that the integrand
contains a product of wavefunctions for a particle- and a hole-state.
The nuclear density at the surface, however, is smaller than for $r=0$
or the saturation density of nuclear matter. Consequently also the
relativistic effects due scalar potential $\Sigma^s$, leading to an
enhacement of the small component of the Dirac spinor, will be smaller
at these relevant densities than at the center of the nucleus or at the
saturation density of nuclear matter. Similar, although a bit smaller,
reductions with respect to the nuclear matter approach were also found
in the finite nuclei perturbative approach of \cite{8}, though the
results were found to be sensitive to the short range correlations
assumed. Here short range correlations are incorporated in the problem
in a selfconsistent way.

{}From these considerations we can also understand that the relativistic
renormalization of the axial charge operator in the case of the
$1p_{3/2} \rightarrow 1d_{3/2}$ transition
is smaller than the one in the $1p_{1/2} \rightarrow 2s_{1/2}$ case.
The smaller renormalization
in the case of the $1p_{3/2} \rightarrow 1d_{3/2}$ transition can be
interpreted in terms of the centrifugal barrier which pushes the $d$ state
more to the surface of the nucleus where the potential $\Sigma^{s}$ is
weaker. Furthermore we observe a slight increase of the renormalization
as a function of $q$. At a larger momentum transfer one tends to probe
more the higher densities in the interior of the nucleus.

These results are confirmed by our calculations for the nucleus $^{40}Ca$.
In fig. 4 we show the relativistic and non relativistic matrix elements for
the $1d_{3/2} \rightarrow 2p_{3/2}$ transition in $^{40}Ca$ and in fig.
5 the
ratio of the relativistic to non relativistic matrix elements.
The ratio is of
the order of 1.23,  rather independent on the momentum transfer. This
result for the renormalization of the axial charge is very similar
but slightly larger than the ratio obtained for the $1p_{3/2} \rightarrow
1d_{3/2}$ in $^{16}O$. Once again the centrifugal barrier is responsible for a
reduced renormalization compared to the expectations of nuclear matter
approach.

\section{Conclusions}

We have analyzed in detail the renormalization of the axial charge in nuclei
by evaluating the matrix elements of the axial charge operator with
relativistic wave functions, solutions of the Dirac equation with the
relativistic potential, and with non relativistic wave function, solutions of
the Schr\"odinger equation with an equivalent non relativistic potential. We
have found renormalization effects due to the use of the relativistic wave
functions, enhancing the axial charge in the direction found in earlier
perturbative approaches for nuclear matter.
However, the quantitative results differ from the estimates derived
for nuclear matter significantly. Using the G-matrix derived from a
realistic meson exchange model of the NN interaction \cite{rupr} a
perturbative estimate of the heavy meson exchange current contribution
to the axial charge at nuclear matter saturation density \cite{7a} would
yield a renormalization factor of 1.4 and a non perturbative  treatment
would lead to enhancement as large as 1.7. For finite nuclei the
enhancement factors considerably smaller, of the order of 1.2 - 1.3.
We argue that this reduction of the renormalization effect is due to
the smaller densities at the surface of finite nuclei, which are
relevant for the evaluation of actual matrix elements. From these
considerations we can also understand the dependence of the
renormalization factor on the momentum transfer and on the transition
actually considered.

The amount of axial charge renormalization depends on the model for the
NN interaction. We have employed a relativistic meson exchange model
(Potential version $A$ of the Bonn potential \cite{rupr}), which has
been derived to reproduce NN scattering data.
It is fair to quote at this point that using this potential in the
present case there is the assumption that the
relativistic potential constructed to reproduce $NN$ scattering of on shell
nucleons can be extrapolated to deal with negative energy states and on shell
and off shell conditions. This is certainly a strong assumption from which all
the microscopically constructed relativistic potentials suffer, and indeed
different parametrizations of the $NN$ amplitude on shell lead to different
relativistic potentials \cite{11}. Some efforts have been done to constrain
the relativistic potential to be consistent with the $\bar{N}N$ elementary
amplitudes \cite{12} and this leads to potentials like the one obtained here
but about one half their strength. Even then this potential is constructed at
the level of the impulse approximation or low density limit, $t{\rho}$, and
many body effects should modify it. It is clear that many efforts are still
necessary to be able to claim that an unambiguous microscopical relativistic
potential has been determined. On the other hand one can take a more
phenomenological approach and say that a certain relativistic potential has a
wide degree of phenomenological success, providing fair nuclear binding
energies, spin-orbit splitting, nucleon nucleus cross sections and
polarization observables, etc. \cite{10,klein}.
  The potential we have used is one of such and provides
empirical support for the axial charge renormalization found,
but this does not
exclude the possibility of other potentials with the same degree of
phenomenological success and still providing different axial charge
renormalization. The ultimate answer to this question is tied to the progress
in our understanding of the meaning and accurate strength of the relativistic
potential. Meanwhile, by using a fair and plausible model we have done
detailed calculations and showed that the results are sufficiently different
from the perturbative results to encourage the use of the present approach in
future works dealing with the problem.

\bigskip\bigskip
\noindent
Two of us, A. Gil and E. Oset wish to acknowledge the hospitality of the
University of T\"ubingen and H. M\"uther the one of the University of Valencia.
E. Oset acknowledges support from the Humboldt Foundation. The work has been
partially supported by the EU, program, Human Capital and Mobility contract
no. CHRX-CT 93-0323,  the CICYT contract no. AEN 93-1205
and the Graduiertenkolleg ``Struktur und Wechselwirkung von Hadronen
und Kernen'' of the Deutsche Forschungsgemeinschaft (DFG Mu 705/3)
\newpage

{\bf Appendix: Matrix elements of the axial charge operator.}

\bigskip\noindent
{\bf A) Relativistic case:}
We write here the matrix element for the $\gamma^{0}
\gamma_{s}$ operator between relativistic wave functions

$$
< n' l' j' m' | \gamma^{0} \gamma_{5} e^{i \vec{q} \vec{r}} | n l j m >
\eqno{(a.1)}
$$

We distinguish 4 cases

$$
\begin{array}{ll}
a) j' = l' + 1/2, & j = l + 1/2\\
b) j' = l' + 1/2, & j = l - 1/2\\
c) j' = l - 1/2, & j = l + 1/2\\
d) j' = l - 1/2, & j = l - 1/2
\end{array}
$$

and the resulting matrix element is

$$
\sqrt{4 \pi} \sum_{\lambda} i^{\lambda} (-i) \int r^{2} dr \left[
\frac{G_{l'j'}(r)}{r} \frac{F_{lj} (r)}{r} - \frac{F_{l'j'}(r)}{r}
\frac{G_{lj} (r)}{r}
\right] j_{\lambda} (qr)
$$

$$
 (2 \lambda + 1)^{1/2} Y_{\lambda , m'-m}^{*} (\hat{q}) A_{i}
\eqno{(a.2)}
$$

\noindent
where $A_{i}$ is given for each of the cases a) b) c) d) listed above by

$$
A_{a}= \frac{1}{2j'} C(j + 1/2, \lambda , j' - 1/2 ; 000)
$$

$$
\{(j' + m')^{1/2} (j + 1 - m)^{1/2} C (j + 1/2, \lambda , j' - 1/2; m -
1/2, m'-m)
$$

$$
-(j' - m')^{1/2} (j + 1 + m)^{1/2} C (j + 1/2, \lambda , j' - 1/2; m+ 1/2,
m' - m) \}
\eqno{(a.3)}
$$

$$
A_{b} = \frac{1}{2j'} C (j-1/2, \lambda , j' - 1/2; 000)
$$

$$
\{ (j' + m')^{1/2} (j + m)^{1/2} C (j - 1/2, \lambda , j'- 1/2; m-1/2, m'-m)
$$

$$
+ (j' - m')^{1/2} (j - m)^{1/2} C (j - 1/2, \lambda , j' - 1/2; m + 1/2, m'-m)
\eqno{(a.4)}
$$

$$
A_{c}= \frac{1}{2j'+2} C (j + 1/2, \lambda , j' + 1/2; 000)
$$

$$
\{ (j' + 1 - m')^{1/2} (j + 1 - m)^{1/2} C (j + 1/2, \lambda , j' + 1/2;
m - 1/2, m' - m)
$$

$$
+ (j' + 1 + m')^{1/2} (j + 1 + m)^{1/2} C (j + 1/2, \lambda , j' + 1/2;
m + 1/2, m' - m)\}
\eqno{(a.5)}
$$

$$
A_{d} = \frac{1}{2j' + 2} C (j - 1/2, \lambda , j' + 1/2; 000)
$$

$$
\{ (j' + 1 - m')^{1/2} (j + m)^{1/2} C(j - 1/2, \lambda , j' + 1/2; m-1/2,
m'-m)
$$

$$
-(j' + 1 + m')^{1/2} (j - m)^{1/2} C (j - 1/2, \lambda , j' + 1/2; m + 1/2,
m'-m) \}
\eqno{(a.6)}
$$
\bigskip\noindent
{\bf B) Non relativistic case:}
we evaluate matrix elements of the $\vec{\sigma}
(\vec{p} + \bar{p}')/2M$ operator between non relativistic states

$$
< n' l' j' m' | \vec{\sigma} (\vec{p} + \vec{p}\:') /2M | n l j m >
\eqno{(a.7)}
$$

\noindent
The derivation of this matrix elements requires a bit more algebra than the
non relativistic case. With the help of some useful formulas from the
appendix of ref. \cite{13} we obtain the following result

$$
< n'j'l'm'| \vec{\sigma} \frac{(\vec{p} + \vec{p}\:')}{2M} e^{i\vec{q}\:
\vec{r}}
| n l j m> =
$$

$$
i(-1)^{j' + l' + 1/2} \sqrt{4 \pi}  \frac{1}{2M} \sum_{\lambda} C(j \lambda
j' ; 1/2,0, 1/2)
$$

$$
C (j \lambda j' ; m, m'-m) i^{\lambda} Y_{\lambda , m'-m}^{*} (\hat{q})
\Bigl[\frac{2 (2
\lambda + 1)}{2j' + 1}\Bigr]^{1/2}
$$

$$
F(n'l'j', n l j ; \lambda)
\eqno{(a.8)}
$$

\noindent
with $\lambda + l + l'$ an odd number, where the last function is given by

$$
F (n' l' j', n l j ; \lambda ) =
$$

$$
 \delta_{j, l + 1/2} (l + 1)^{1/2} \int_{0}^{\infty} r^{2} dr \phi_{l'} (r)
[\frac{d \phi_{l}(r)}{dr} - \frac{l}{r} \phi_{l}(r)] j_{\lambda} (qr)
$$

$$
- (-1)^{j-j' +l-l'} \delta_{j', l' + 1/2} (l'+1)^{1/2} \left( \frac{2j + 1}
{2j' + 1} \right)^{1/2}
$$

$$
\int_{0}^{\infty} r^{2} dr [\frac{d \phi_{ l'}(r)}{dr} - \frac{l'}{r}
\phi_{l'}(r)]
\phi_{l}(r)j_{\lambda} (qr)
$$

$$
- \delta _{j, l- 1/2} l^{1/2} \int_{0}^{\infty} r^{2} dr \phi_{l'}(r) [
\frac{d \phi_{l}(r)}{dr} + \frac{l + 1}{r} \phi_{l}(r)] j_{\lambda} (qr)
$$

$$
+ (-1)^{j-j'+l -l'} \delta_{j', l' - 1/2} l'^{1/2}
\left(\frac{2j + 1}{2j' + 1}\right)^{1/2}
$$

$$
\int_{0}^{\infty} r^{2} dr [\frac{d \phi_{l'}(r)}{dr} +
\frac{l' + 1}{r} \phi_{l'}(r)]
\phi_{l} (r) j_{\lambda} (qr)
\eqno{(a.9)}
$$

\newpage

\noindent{\bf Figure captions:}

\bigskip\noindent
Fig. 1 Diagrams appearing in the perturbative approach to the renormalization
of the axial charge.

a) b) involving positive energy intermediate nucleon states: c) d) involving
negative energy intermediate states.

\bigskip\noindent
Fig. 2 Axial charge form factor for the $0^{+} \rightarrow 0^{-}$ transition
in $^{16}O$ from the orbitals $1p_{1/2} \rightarrow 2s_{1/2}$ and
$1p_{3/2} \rightarrow 1d_{3/2}$ with relativistic and equivalent non
relativistic wave functions.

\bigskip\noindent
Fig. 3 Ratio of the relativistic to non relativistic matrix elements of fig.
2 as a function of the momentum transfer.

\bigskip\noindent
Fig. 4 Same as fig. 3 for $^{40}Ca$ and the transition $0^{+}\rightarrow
0^{-}$ from the
orbitals $1d_{3/2} \rightarrow 2p_{3/2}$.

\bigskip\noindent
Fig. 5 Ratio of relativistic to non relativistic matrix elements for the
transition in fig. 4.


\newpage

\begin{thebibliography}{99}
\bibitem{1} J. Delorme and I.S. Towner, Nucl. Phys. A475 (1987) 720
\bibitem{2} B.D. Serot and J.D. Walecka, Adv. Nucl. Phys. 16 (1986) 1 ;\\
J.D. Walecka, Ann of Phys. 83 (1974) 491.
\bibitem{3} M. Kubodera, J. Delorme and M. Rho, Phys. Rev. Lett. 40 (1978) 755
\bibitem{4} J. Delorme, Nucl. Phys. A374 (1982) 541.
\bibitem{5} M. Kirchbach, S. Kamalov and H.U. Jager,
Phys. Lett. B144 (1984) 319;\\
H.U. Jager, M. Kirchbach and E. Truhlik, Nucl. Phys. A404 (1983) 456.
\bibitem{6} M. Kirchbach and H. Reinhard, Phys. Lett. B208 (1988) 79.
\bibitem{7} M. Kirchbach, D. O. Riska and K. Tsushima, Nucl. Phys.
A542 (1992) 616.
\bibitem{7a} M. Hjorth-Jensen, M. Kirchbach, D.O. Riska snd K.
Tsushima, Nucl. Phys. A 563 (1993) 525.
\bibitem{8} I.S. Towner, Nucl. Phys. A542 (1992) 631.
\bibitem{10} R. Fritz, H. M\"uther and R. Machleidt, Phys. Rev. Lett. 71
(1993) 46; \\
R. Fritz and H. M\"uther, Phys. Rev. C 49 (1994) 633.
\bibitem{rupr} R.~Machleidt, Adv. in Nucl. Phys.  19 (1989) 189.
\bibitem{klein} M.~Kleinmann, R.~Fritz, H.~M\"uther and A.~Ramos, Nucl.
Phys. in print
\bibitem{9} F. Mandl and G. Shaw, Quantum Field Theory, John Wiley, 1988.
\bibitem{itz} C. Itzykson and J.B. Zuber, ``Quantum Field Theory''
(McGraw-Hill, New York 1980)
\bibitem{jam89} M.~Jaminon and C.~Mahaux, Rev. {\bf C 40} (1989) 354.
\bibitem{11} C.J. Horowitz, Phys. rev. C31 (1985) 1340;\\
L. Ray, G.W. Hoffmann and W.R. Coker, Phys. Reports 212 (1992) 223.
\bibitem{12} J.A. Tjon and S.J. Wallace, Phys. Rev. C32 (1985) 1667;\\
ibid C35 (1987) 280;\\
ibid C36 (1987) 1085.
\bibitem{13} A. Galindo and P. Pascual, Quantum Mechanics, Springer, Berlin
1991.
\end{thebibliography}
\end{document}